\documentclass[fleqn]{article}
\usepackage{longtable}
\usepackage{graphicx}

\begin {document}
\begin{flushleft}
{\LARGE
{\bf Comment on ``Multiconfiguration Dirac-Fock energy levels and radiative rates for Br-like tungsten" by \\S. Aggarwal, A.K.S. Jha, and M. Mohan [Can . J. Phys. 91 (2013) 394]}
}\\

\vspace{1.5 cm}

{\bf {Kanti  M  ~Aggarwal and  Francis   P   ~Keenan}}\\ 

\vspace*{1.0cm}

Astrophysics Research Centre, School of Mathematics and Physics, Queen's University Belfast, Belfast BT7 1NN, Northern Ireland, UK\\ 
\vspace*{0.5 cm} 

e-mail: K.Aggarwal@qub.ac.uk \\

\vspace*{0.20cm}

Received  22 July 2013.  Accepted 23 October 2013 \\

\vspace*{1.5cm}

PACS Ref: 31.25 Jf, 32.70 Cs,  95.30 Ky

\vspace*{1.0 cm}

\hrule

\vspace{0.5 cm}

\end{flushleft}

\clearpage


\begin{abstract}

We report  calculations of energy levels and oscillator strengths for transitions in W XL, undertaken with the general-purpose relativistic atomic structure package ({\sc grasp}) and flexible atomic code ({\sc fac}). Comparisons are made with existing results and the accuracy of the data is assessed. Discrepancies with the most recent results of S. Aggarwal et al. [Can. J. Phys. {\bf 91} (2013) 394] are up to 0.4 Ryd  and up to two orders of magnitude for energy levels and oscillator strengths, respectively.  Discrepancies for lifetimes are even larger, up to four orders of magnitude for some levels. Our energy levels are estimated to be accurate to better than 0.5\% (i.e.  0.2 Ryd), whereas results for oscillator strengths and lifetimes should be  accurate to  better than 20\%.

\end{abstract}

\clearpage

\section{Introduction}

Tungsten (W) is an important constituent of tokamak reactor walls, and hence to study fusion plasmas, especially to assess  radiation loss, atomic data (including energy levels and  oscillator strengths or radiative decay rates) are required for many of its ions. The need for atomic data has become even greater with the developing  ITER project. Considering its importance, there have been several theoretical studies for W ions -- see for example, Fournier \cite{kbf}. Similarly, there have been several line measurements   -- see for example, Utter et al. \cite{sbu} and Clementson et al. \cite{ll1}.  Laboratory measurements for W emission lines, including W XL,  have been  compiled by NIST (National Institute of Standards and Technology), and are  available at their  website {\tt http://physics.nist.gov/PhysRefData/ASD/levels\_form.html}.

Recently, S. Aggarwal et al. (\cite{ajm}, henceforth to be referred to as AJM) have reported results for energy levels, oscillator strengths, radiative rates,  and lifetimes for Br-like W XL. For their calculations, they adopted the {\sc grasp} (general-purpose relativistic atomic structure package) code to generate the wavefunctions. This code was originally developed as GRASP0 by Grant  et al.  \cite{grasp0} and has been updated by Dr. P. H. Norrington ({\tt http://web.am.qub.ac.uk/DARC/}). It is a fully relativistic code and is based on the $jj$ coupling scheme. Further relativistic corrections arising from the Breit interaction and QED (quantum electrodynamics) effects have also been included. 

For a heavy ion such as W XL, relativistic effects are very important for an accurate determination of energy levels, and subsequently other  parameters, including radiative rates and lifetimes. However, for this ion {\em configuration interaction} (CI) is also very important, because levels of many of its configurations closely interact and intermix. For this reason,  AJM {\cite{ajm} included CI among 21 configurations, but  their results show discrepancies with the NIST listings of up to 0.8 Ryd. In addition,  their calculations differ from those of Fournier \cite{kbf} by up to 0.5 Ryd for several levels (see Table 2). This is in spite of the fact that Fournier has also included a comparable large CI, among 20 configurations.  More importantly, oscillator strengths for some transitions differ by up to two orders of magnitude (see Table 3), which subsequently affect the calculations of lifetimes. The main reason for these discrepancies is that both workers considered different sets of configurations (see Table 1). For example, AJM have included the 4p$^3$4d$^2$ configuration whereas Fournier has not. This configuration generates 141 odd parity levels in the 22--42 Ryd energy range. Similarly, Fournier has included the 3d$^9$4s$^2$4p$^5$4d/4f and 3p$^5$3d$^{10}$4s$^2$4p$^5$4d/4f configurations, but AJM have not. Although these configurations generate levels  in the higher  energy range above 130 Ryd, they also affect the calculations for many levels of other configurations, especially  3d$^9$4s$^2$4p$^6$ (120--126 Ryd), for which both workers have reported results. Finally, both workers have {\em omitted} some of the important configurations, such as 4s$^2$4p$^3$4d4f and 4s$^2$4p$^2$4d$^3$, which together generate 624 levels  in the 34--58 Ryd energy range.

Although there is a need for atomic data for tungsten ions as stated above, these must also be {\em reliable} (see for example Aggarwal and Keenan \cite{fst}), so they  can be confidently applied to the modelling of plasmas. Therefore, our {\em aim} in this short communication is to improve upon the accuracy of atomic data for W XL, and to resolve discrepancies with earlier calculations. 


\section{Energy levels}

For our calculations we have adopted the same  {\sc grasp}  code as employed by AJM {\cite{ajm}. Similarly, we have also used the option of {\em extended average level} (EAL),  in which a weighted (proportional to 2$j$+1) trace of the Hamiltonian matrix is minimised. This produces a compromise set of orbitals describing closely-lying states with moderate accuracy, and generally yields results comparable to other options, such as {\em average level} (AL), as noted by Aggarwal  et al.  for several ions of Kr \cite{kr} and Xe \cite{xe}.  
Furthermore, to  assess the accuracy of our results, we have also employed the  {\em Flexible Atomic Code} ({\sc fac}) of Gu \cite{fac},  available from the website {\tt http://sprg.ssl.berkeley.edu/$\sim$mfgu/fac/}. This is also a fully relativistic code which provides a variety of atomic parameters, and yields results for energy levels and oscillator strengths comparable to {\sc grasp}, as already shown for several other ions, see for example:  Aggarwal  et al. for Kr \cite{kr} and Xe \cite{xe} ions. Additionally, very large calculations can  be performed with this code and within a reasonable time frame of a few days. Thus results from {\sc fac} will be helpful in assessing the accuracy of our energy levels and radiative rates.

Since CI is very important for W XL as noted above, we have performed a series of calculations  with the {\sc grasp} code with increasing amount of CI,  but focus only on three, namely (i) GRASP1, which includes 63 levels among the 4s$^2$4p$^5$, 4s$^2$4p$^4$4d, 4s$^2$4p$^4$4f,  4s4p$^6$, and 3d$^9$4s$^2$4p$^6$ configurations, (ii) GRASP2, with a total of 3490 levels, the additional ones arising from the 17   configurations  4p$^6$4d/4f, 4s4p$^5$4d/4f, 4p$^3$4d$^2$/4f$^2$/4d4f, 4s$^2$4p$^2$4d$^3$, 4s$^2$4p4d$^4$, 4s$^2$4p$^2$4d$^2$4f, 4s4p$^3$4d$^3$, 4p$^5$4d$^2$, 3d$^9$4s$^2$4p$^5$4d/4f, 3p$^5$3d$^{10}$4s$^2$4p$^6$,  and 3p$^5$3d$^{10}$4s$^2$4p$^5$4d/4f, and finally (iii) GRASP3, which includes levels from a further 20 configurations, namely 4s4p$^5$5$\ell$, 4p$^6$5$\ell$, 4s$^2$4p$^4$5$\ell$, and 3d$^9$4s$^2$4p$^5$5$\ell$, i.e. 4128 levels in total among all  42 configurations. These  are  listed in Table 1 along with the number of levels each configuration generates and their energy ranges. Furthermore, to facilitate  comparison, the configurations included by Fournier \cite{kbf} and AJM {\cite{ajm} are also marked.

As with {\sc grasp}, we have also performed a series of calculations  with the {\sc fac} code with increasing amount of CI,  but focus only on two. These are (i) FAC1, which includes the same 4128 levels as in GRASP3, and (ii) FAC2, which also includes all possible combinations of the 4s$^2$4p$^3$5$\ell{\ell'}$ and 4s4p$^4$5$\ell{\ell'}$  configurations, i.e. 11,525 levels in total. Results obtained from all these five calculations are listed in Table 2, along with those of  NIST, Fournier \cite{kbf} and AJM {\cite{ajm}. However, data  are provided here for only 33 levels, common to the earlier calculations,  belonging to the 4s$^2$4p$^5$, 4s$^2$4p$^4$4d, 4s4p$^6$, and 3d$^9$4s$^2$4p$^6$ configurations. Before we undertake comparisons, we note that NIST listings are not very accurate for some levels. For example, the quoted uncertainties for levels (4s$^2$4p$^4$4d) $^4$F$_{7/2}$ and $^2$P$_{1/2}$ (6 and 9) are 5000 cm$^{-1}$, or 0.0456 Ryd. Similarly, the NIST listings for levels 23--25 are the same, and hence are not accurate. Furthermore, NIST energies are not available for all the desired levels -- see Table 2.

Some of the levels are highly mixed and this has also been noted by AJM {\cite{ajm} -- see their Table 1. Therefore, it is not always possible to provide a unique label for each level, but care has been taken to identify the levels as accurately as possible. However, the best one can say about a level is that it has a particular  $J$ value, as listed in Table 2, but there can be disagreements about the configuration assigned to it. Based on several calculations and our past experience for a wide range of ions, we have assigned a configuration for each level, and the only differences with the listings of AJM  are for levels 13 and 26, which are interchanged, i.e. (4s$^2$4p$^4$4d) $^2$F$_{5/2}$ and $^2$D$_{5/2}$. Finally, we stress that calculations performed with the {\em same} configurations as adopted by AJM  yield comparable results (for both energy levels and oscillator strengths) as reported by them and listed under the  column GRASP4  in Tables 2 and 3. Therefore, discrepancies between our other calculations and  their results are only because of the different  configurations included.

Our results from GRASP1 are closest to those of NIST, and the maximum discrepancy of $\sim$ 0.3 Ryd is for  levels 23--25, and for the last  (3d$^9$4s$^2$4p$^6$ $^2$D$_{3/2}$) our energy is higher by 0.4 Ryd. The GRASP2 calculations include larger CI and hence  are comparatively more accurate, but discrepancies with GRASP1 are up to 0.2 Ryd, execpt for the last two levels for which the differences are up to 0.4 Ryd. Furthermore,  for a majority of levels the GRASP2 energies are higher than those from GRASP1, and for this reason discrepancies with the NIST listings are up to 0.7 Ryd, particularly for  levels of the 3d$^9$4s$^2$4p$^6$ configuration (32--33). The inclusion of the 5$\ell$ configurations in the GRASP3 calculations has an insignificant effect on the energies  obtained with  GRASP2, because the differences (if any) are within 0.1 Ryd. For the lowest 31 levels, the energies obtained with FAC1 agree closely (within 0.07 Ryd) with GRASP3, but differences are slightly larger (up to  0.2 Ryd) for the last two levels. Since both the GRASP3 and FAC1 results include the same CI, the two sets of energies are in close agreement. Inclusion of larger CI in FAC2 does not appreciably improve the accuracy of the energy levels, because differences with FAC1 are within 0.01 Ryd. Therefore, we may conclude that CI among the $n$=4 configurations is more important (and nearly sufficient) than with $n$=5.

The differences in energies between our calculations (GRASP3 and FAC1) and those of Fournier \cite{kbf} are up to 0.4 Ryd for several levels -- see for example, the lowest 10, 14, and the last two. This is mainly because he has not included several important configurations, as shown in Table 1. The energies reported by AJM {\cite{ajm} also differ from our calculations, by up to 0.4 Ryd, for several levels, such as 23 and above, and their results are mostly higher. This is clearly due to the limited CI included by them as stated earlier and demonstrated in Tables 1 and 2. Overall, we may state with confidence that  our GRASP3 and/or FAC1/FAC2 results of energy levels for W XL  in Table 2 are the most accurate available to date.

\section{Radiative rates}

The absorption oscillator strength ($f_{ij}$), a dimensionless quantity,  and radiative rate A$_{ji}$ (in s$^{-1}$) for a transition $i \to j$ are related by the following expression:

\begin{equation}
f_{ij} = \frac{mc}{8{\pi}^2{e^2}}{\lambda^2_{ji}} \frac{{\omega}_j}{{\omega}_i}A_{ji}
 = 1.49 \times 10^{-16} \lambda^2_{ji} (\omega_j/\omega_i) A_{ji}
\end{equation}
where $m$ and $e$ are the electron mass and charge, respectively, $c$  the velocity of light,  $\lambda_{ji}$  the transition energy/wavelength in $\rm \AA$, and $\omega_i$ and $\omega_j$  the statistical weights of the lower $i$ and upper $j$ levels, respectively.

In Table 3 we compare our f- values from three calculations with {\sc grasp} (GRASP1, GRASP2, and GRASP3) and two with {\sc fac} (FAC1 and FAC2), with those of Fournier \cite{kbf} and AJM {\cite{ajm}. The A- and f- values have been calculated in both Babushkin and Coulomb gauges, which are  equivalent to the length and velocity forms in the non-relativistic nomenclature. However, as for earlier calculations,  data are presented here in the length form alone, because these are considered to be comparatively more accurate.  The  results obtained with GRASP2 for the f- values are comparable with those from GRASP3, for most of the transitions (as also found for energy levels), but the differences for a few weak transitions (f $\sim$ 10$^{-5}$), particularly 1--30 and 2--20, are up to a factor of two. However, the f- values for some  transitions from GRASP1 differ from GRASP2 and GRASP3 by up to three orders of magnitude, see for example, 1--11/23 and 2--12/20. This is clearly due to the limited CI included in the GRASP1 calculations. On the other hand, results obtained from GRASP3 and FAC1  are comparable for almost all transitions,  the only exception being  1--23 and 1--29 (f  $\sim$ 10$^{-2}$ and 10$^{-4}$, respectively) for which the two results differ by up to $\sim$50\%. In general, both codes with comparable CI yield similar results for a majority of transitions, and hence support the reliability of our calculations. Furthermore, the additional CI included in FAC2 is of no clear advantage, because the f- values agree within a few percent with those from FAC1. This conclusion is consistent with that  for the energy levels in section 2.

The f- values of Fournier \cite{kbf} agree with our calculations with {\sc grasp} and {\sc fac} to within a factor of three for all transitions listed in Table 3. However, the discrepancies with the other calculations of AJM {\cite{ajm} are unfortunately up to two orders of magnitude for some  transitions, see for example, 1--22/26 and 2--20/25. The AJM f-values show similar differences  with the calculations of  Fournier, because they have omitted some of the crucial configurations (see Table 1). Based on the comparisons made in Table 3 and discussed above, we may state  that the f- values reported by AJM  {\cite{ajm} are not reliable. On the other hand, based on the consistency of results between our {\sc grasp} and {\sc fac} calculations, we may conclude with confidence that our f- values listed in Table 3 are accurate to better than 20\%, for a majority of transitions.

Apart from the convergence and consistency tests discussed above, one of the  criteria normally used to assess the accuracy of f- (or A-) values is the agreement between the Babushkin and Coulomb gauges, or equivalently the length and velocity forms. Before we discuss this, we note that a good agreement between the two forms is only a desirable but not a necessary condition. This is because different sets of configurations may result in equally good agreement between the two forms, but entirely different results in magnitude, not only for the weak(er) transitions, but also the allowed ones which are comparatively greater  and more stable in magnitude. Examples of this can be seen in \cite{ah1}--\cite{kma}. Nevertheless, in Table 3 we list the velocity/length ratio of the f- values corresponding only to our GRASP3 and GRASP4 calculations, i.e R3 and R4, respectively. Almost for all transitions, strong as well as weak, based on the ratio alone the f- values corresponding to the GRASP4 calculations should be comparatively more accurate. However, this is not the case as discussed above and f- values obtained in the GRASP2/3 and FAC1/2 calculations are not only consistent but also more accurate. Finally, based on the ratio R3 the accuracy of our listed results is better than 20\% for a majority of the transitions as stated above.

\section{Lifetimes}

The lifetime $\tau$ of a level $j$ is defined as follows:

\begin{equation}
{\tau}_j = \frac{1}{{\sum_{i}^{}} A_{ji}}.
\end{equation}

In Table 4 we list lifetimes for all 33 levels from our calculations with the {\sc grasp} code. These results correspond to the GRASP3 calculations and {\em include} A- values from all types of transitions, i.e. E1 (electric dipole), E2 (electric quadrupole), M1 (magnetic dipole), and M2 (magnetic quadrupole). Unfortunately, there are no measurements   available with which to compare our results. However, AJM {\cite{ajm} have reported lifetimes for these levels, which are included in Table 4 for comparison. As for our lifetimes,  they too have included contributions from all four types of transition.

The discrepancies between our  calculated lifetimes and those of AJM {\cite{ajm} are up to four orders of magnitude for some levels, such as 10, 18, 27, and 30. Although our (GRASP3) calculations include a larger CI, considering that both calculations have adopted the same {\sc grasp} code such large discrepancies are unexpected and puzzling. As noted above in section 2, the energy levels and f-  (or A- values) obtained by us with the same configuration set as employed by  AJM (GRASP4) are comparable with their results. Therefore, to understand the differences, in Table 4 we have also listed lifetimes obtained  from our calculation which adopts the AJM configuration set, which we designate as GRASP4a. Additionally, we have also listed the A- values for the {\em dominant} transitions. It is satisfying to note that there is {\em no} major discrepancy between the GRASP3 and GRASP4a lifetimes for most of the levels. However, for some  the differences are up to a factor of five, and examples include 11, 20, and 24. These discrepancies are easily understandable, because they correspond to the similar differences in the f- (or A-) values, as seen in Table 3. 

The discrepancies between the lifetimes of  AJM {\cite{ajm}, designated GRASP4b in Table 4,  and our GRASP4a results are even larger than with GRASP3 -- see level 30. Since the discrepancies are the largest for two levels, namely (4s$^2$4p$^4$($^3$P)4d) $^4$F$_{9/2}$ and $^2$D$_{3/2}$ (10 and 30), we focus our efforts on these two. For level 10, apart from the dominant 6--10 M1 transition (A = 2.68$\times$10$^4$ s$^{-1}$), the only other contributing transitions are 4--10 E2 (A = 3.57 s$^{-1}$) and 6--10 E2 (A = 7.53 s$^{-1}$).  These latter two yield a total of  A = 11.1  s$^{-1}$, or equivalently, $\tau$ = 0.09  s, a value close to the 0.127  s reported by AJM. Therefore, it appears that they have not included the contribution of the 6--10 M1 transition. For level 30, the dominant contribution to $\tau$ is of the 2--30 E1 (A = 1.81 $\times$10$^{12}$ s$^{-1}$) transition, for which there is no discrepancy among the f- (or A-) values in GRASP3, GRASP4a and GRASP4b calculations as already shown in Table 3. Therefore, the $\tau$ value of 8.66$\times$10$^{-8}$ s reported by AJM is unrealistic. However, the next important contribution for this level is of the 1--30 E1 (A = 1.16 $\times$10$^{7}$ s$^{-1}$) transition, which yields $\tau$ = 8.61$\times$10$^{-8}$  s, a value closer to that reported by AJM. Therefore, their lifetime for this level appears to be based on the 1--30 E1 transition alone. In conclusion,  the large discrepancies for several levels indicate the unreliability of the lifetimes reported by  AJM {\cite{ajm}.

\section{Conclusions}

In this work, energy levels, radiative rates, oscillator strengths, and lifetimes have been calculated for a large number of levels/transitions of W XL. For the calculations, two independent codes, {\sc grasp} and {\sc fac}, have been adopted in order to rigorously assess the accuracy of the desired atomic data. However, results are presented only for 33 levels of the 4s$^2$4p$^5$, 4s$^2$4p$^4$4d, 4s4p$^6$, and 3d$^9$4s$^2$4p$^6$ configurations, mainly because similar calculations and for the same levels have recently been reported by AJM {\cite{ajm}. Results for a larger number of levels and their corresponding transitions  will be reported in a later paper.

For all the levels, there are no discrepancies in their energies among our calculations, but differences with the results of AJM \cite{ajm} are up to 0.4 Ryd for some levels. Discrepancies with their f- values are even greater, up to two orders of magnitude, for some of the transitions, due to the limited CI included by them. Finally, their reported lifetimes show some large errors, of up to four orders of magnitude, for several levels, and hence are unreliable. 

Based on a variety of comparisons among different calculations, as well as with the earlier work of Fournier \cite{kbf} and the NIST compilations of experimental energies, our reported energy levels are assessed to be accurate to better than 0.5\%. Similarly, the accuracy for other parameters, namely oscillator strengths and lifetimes, is assessed to be better than 20\%.

\section*{Acknowledgment}
 KMA  is thankful to  AWE Aldermaston for financial support.     



\newpage
\clearpage

\begin{flushleft}
Table 1. Configurations and  levels of W XL. 
\end{flushleft}
\begin{tabular}{rllccccccrrrr} \hline
\\
Index  & Configuration      & No. of Levels  & Energy Range (Ryd)    &  GRASP1 & GRASP2  & GRASP3  & GRASP4  &  RELAC     \\
\\ \hline
  1  &  4s$^2$4p$^5$ 		        &     2$^o$  & 0--7	& Y & Y & Y & Y  & Y \\
  2  &  4s$^2$4p$^4$4d 		        &    28      & 11--27	& Y & Y & Y & Y  & Y \\
  3  &  4s$^2$4p$^4$4f 		        &    30$^o$  & 25--42	& Y & Y & Y & Y  & Y \\
  4  &  4s4p$^6$		        &     1      & 15	& Y & Y & Y & Y  & Y \\
  5  &  4p$^6$4d		        &     2      & 43--45	&   & Y & Y & Y  &   \\
  6  &  4p$^6$4f		        &     2$^o$  & 58--59	&   & Y & Y &	 &   \\
  7  &  4s4p$^5$4d		        &    23$^o$  & 25--36	&   & Y & Y & Y  & Y \\
  8  &  4s4p$^5$4f		        &    24      & 41--50	&   & Y & Y & Y  & Y \\
  9  &  4s$^2$4p$^3$4d$^2$		&   141$^o$  & 22--42	&   & Y & Y & Y  &   \\
 10  &  4s$^2$4p$^3$4f$^2$		&   221$^o$  & 52--71	&   & Y & Y & Y  &   \\
 11  &  4s$^2$4p$^3$4d4f		&   363      & 37--58	&   & Y & Y &	 &   \\
 12  &  4s$^2$4p$^2$4d$^3$		&   261      & 34--57	&   & Y & Y &	 &   \\
 13  &  4s$^2$4p4d$^4$		        &   180$^o$  & 54--73	&   & Y & Y &	 &   \\
 14  &  4s$^2$4p$^2$4d$^2$4f		&  1140$^o$  & 49--73	&   & Y & Y &	 &   \\
 15  &  4s4p$^3$4d$^3$		        &   678$^o$  & 47--73	&   & Y & Y &	 &   \\
 16  &  4p$^5$4d$^2$		        &    45$^o$  & 53--64	&   & Y & Y &	 &   \\
 17  &  3d$^9$4s$^2$4p$^5$4d	        &    96$^o$  & 131--147 &   & Y & Y &	 & Y \\
 18  &  3d$^9$4s$^2$4p$^5$4f	        &   113      & 147--161 &   & Y & Y &	 & Y \\
 19  &  3d$^9$4s$^2$4p$^6$	        &     2      & 120--126 & Y & Y & Y & Y  & Y \\
 20  &  3p$^5$3d$^{10}$4s$^2$4p$^6$     &     2$^o$  & 155--178 &   & Y & Y &	 & Y \\
 21  &  3p$^5$3d$^{10}$4s$^2$4p$^5$4d   &    65      & 165--198 &   & Y & Y &	 & Y \\
 22  &  3p$^5$3d$^{10}$4s$^2$4p$^5$4f   &    36$^o$  & 180--212 &   & Y & Y &	 & Y \\
 23  &  4s4p$^5$5s		        &     7$^o$  & 65--74	&   &	& Y & Y  &   \\
 24  &  4s4p$^5$5p		        &    18      & 68--80	&   &	& Y & Y  &   \\
 25  &  4s4p$^5$5d		        &    23$^o$  & 77--86	&   &	& Y & Y  &   \\
 26  &  4s4p$^5$5f		        &    24      & 84--92	&   &	& Y &	 &   \\
 27  &  4s4p$^5$5g		        &    24$^o$  & 87--96	&   &	& Y &	 &   \\
 28  &  4p$^6$5s		        &     1      & 82	&   &	& Y &	 &   \\
 29  &  4p$^6$5p		        &     2$^o$  & 86--89	&   &	& Y &	 &   \\
 30  &  4p$^6$5d		        &     2      & 94--95	&   &	& Y &	 &   \\
 31  &  4p$^6$5f		        &     2$^o$  & 100--101 &   &	& Y &	 &   \\
 32  &  4p$^6$5g		        &     2      & 104--105 &   &	& Y &	 &   \\
 33  &  4s$^2$4p$^4$5s		        &     8      & 50--65	&   &	& Y & Y  & Y \\
 34  &  4s$^2$4p$^4$5p		        &    21$^o$  & 54--72	&   &	& Y & Y  & Y \\
 35  &  4s$^2$4p$^4$5d		        &    28      & 62--78	&   &	& Y & Y  & Y \\
 36  &  4s$^2$4p$^4$5f		        &    30$^o$  & 69--84	&   &	& Y & Y  & Y \\
 37  &  4s$^2$4p$^4$5g		        &    30      & 73--88	&   &	& Y &	 &   \\
 38  &  3d$^9$4s$^2$4p$^5$5s	        &    23$^o$  & 172--185 &   &	& Y & Y  & Y \\
 39  &  3d$^9$4s$^2$4p$^5$5p	        &    65      & 175--191 &   &	& Y & Y  & Y \\
 40  &  3d$^9$4s$^2$4p$^5$5d	        &    96$^o$  & 183--197 &   &	& Y & Y  & Y \\
 41  &  3d$^9$4s$^2$4p$^5$5f	        &   113      & 190--203 &   &	& Y & Y  & Y \\
 42  &  3d$^9$4s$^2$4p$^5$5g	        &   119$^o$  & 194--207 &   &	& Y &	 &   \\
 \\ \hline  											      
\end{tabular}   								   					       
			      							   					       
\vspace*{0.5 cm}													       
\begin{flushleft}													       
{\small
GRASP1: present calculations from the {\sc grasp} code with 63 levels \\ 
GRASP2: present calculations from the {\sc grasp} code with 3490 levels\\
GRASP3: present calculations from the {\sc grasp} code with 4128 levels\\ 
GRASP4: AJM  \cite{ajm}, 638 levels among 21 configurations \\ 
RELAC: Fournier \cite{kbf} \\ 
Y: configuration included in the calculation \\										       
															       
}															       
\end{flushleft} 

\newpage
\clearpage
\begin{flushleft}
Table 2. Energies (Ryd) for some levels of W XL. 
\end{flushleft}
\begin{tabular}{rllrrrrrrrrrr} \hline
 & & & & & & & &  \\
Index  & Configuration      & Level              & NIST     &  GRASP1 & GRASP2  & GRASP3  & FAC1    & FAC2	&  RELAC   & GRASP4    \\
& & & & & & & &   \\ \hline
& & & & & & & &   \\
  1  &  4s$^2$4p$^5$	   	   &  $^2$P$^o_{3/2}$	&  0.0000  &  0.0000 &  0.0000 &  0.0000 &  0.0000 &   0.0000  &  0.0000  &  0.0000  \\
  2  &  4s$^2$4p$^5$	   	   &  $^2$P$^o_{1/2}$	&  6.7632  &  6.8419 &  6.7972 &  6.7980 &  6.8104 &   6.8102  &  6.8213  &  6.8004  \\
  3  &  4s$^2$4p$^4$($^3$P)4d	   &  $^4$D$  _{3/2}$	&	   & 11.0344 & 11.2371 & 11.2569 & 11.2699 &  11.2780  & 11.0971  &  11.4291 \\
  4  &  4s$^2$4p$^4$($^3$P)4d	   &  $^4$D$  _{5/2}$	&	   & 11.2042 & 11.4047 & 11.4241 & 11.4371 &  11.4450  & 11.2652  &  11.5956 \\
  5  &  4s$^2$4p$^4$($^3$P)4d	   &  $^4$P$  _{1/2}$	& 11.2290  & 11.2068 & 11.4082 & 11.4280 & 11.4398 &  11.4475  & 11.2788  &  11.6140 \\
  6  &  4s$^2$4p$^4$($^3$P)4d	   &  $^4$F$  _{7/2}$	& 11.4100  & 11.4084 & 11.5931 & 11.6105 & 11.6220 &  11.6296  & 11.4609  &  11.7900 \\
  7  &  4s$^2$4p$^4$($^1$S)4d	   &  $^2$D$  _{3/2}$	&	   & 12.0304 & 12.1315 & 12.1503 & 12.1597 &  12.1673  & 12.0890  &  12.3756 \\
  8  &  4s$^2$4p$^4$($^3$P)4d	   &  $^2$F$  _{7/2}$	& 12.5338  & 12.5350 & 12.7295 & 12.7466 & 12.7644 &  12.7724  & 12.5901  &  12.9186 \\
  9  &  4s$^2$4p$^4$($^3$P)4d	   &  $^2$P$  _{1/2}$	& 12.5852  & 12.5745 & 12.7966 & 12.8145 & 12.8296 &  12.8367  & 12.6535  &  12.9829 \\
 10  &  4s$^2$4p$^4$($^3$P)4d	   &  $^4$F$  _{9/2}$	&	   & 12.6163 & 12.7961 & 12.8113 & 12.8290 &  12.8369  & 12.6604  &  12.9872 \\
 11  &  4s$^2$4p$^4$($^3$P)4d	   &  $^4$P$  _{5/2}$	&	   & 13.6091 & 13.7229 & 13.7413 & 13.7548 &  13.7621  & 13.6743  &  13.9661 \\
 12  &  4s$^2$4p$^4$($^3$P)4d	   &  $^4$P$  _{3/2}$	& 13.8791  & 13.9843 & 14.0765 & 14.0922 & 14.1067 &  14.1136  & 14.0390  &  14.3881 \\
 13  &  4s$^2$4p$^4$($^1$D)4d	   &  $^2$F$  _{5/2}$	& 14.0930  & 14.2332 & 14.3116 & 14.3243 & 14.3387 &  14.3457  & 14.2659  &  14.6104 \\
 14  &  4s4p$^6$                   &  $^2$S$  _{1/2}$	& 14.9266  & 15.0298 & 15.3623 & 15.3667 & 15.3781 &  15.3795  & 15.0905  &  15.4371 \\
 15  &  4s$^2$4p$^4$($^3$P)4d	   &  $^4$D$  _{1/2}$	&	   & 17.4738 & 17.6537 & 17.6703 & 17.7109 &  17.7190  &	  &  17.8556 \\
 16  &  4s$^2$4p$^4$($^3$P)4d	   &  $^4$F$  _{3/2}$	&	   & 17.8421 & 18.0148 & 18.0306 & 18.0697 &  18.0775  &	  &  18.2259 \\
 17  &  4s$^2$4p$^4$($^3$P)4d	   &  $^4$F$  _{5/2}$	&	   & 18.1725 & 18.3227 & 18.3367 & 18.3737 &  18.3812  & 18.2216  &  18.5495 \\
 18  &  4s$^2$4p$^4$($^1$D)4d	   &  $^2$G$  _{7/2}$	&	   & 18.2514 & 18.3940 & 18.4065 & 18.4433 &  18.4508  &	  &  18.6177 \\
 19  &  4s$^2$4p$^4$($^3$P)4d	   &  $^4$D$  _{7/2}$	&	   & 19.1473 & 19.3205 & 19.3343 & 19.3790 &  19.3869  &	  &  19.5256 \\
 20  &  4s$^2$4p$^4$($^1$D)4d	   &  $^2$P$  _{3/2}$	&	   & 19.4313 & 19.5844 & 19.5979 & 19.6372 &  19.6443  & 19.5063  &  19.8387 \\
 21  &  4s$^2$4p$^4$($^1$D)4d	   &  $^2$G$  _{9/2}$	&	   & 19.6369 & 19.7835 & 19.7948 & 19.8376 &  19.8007  & 19.6807  &  20.0002 \\
 22  &  4s$^2$4p$^4$($^3$P)4d	   &  $^2$F$  _{5/2}$	&	   & 19.6180 & 19.7432 & 19.7507 & 19.7843 &  19.7906  & 19.6824  &  20.0090 \\
 23  &  4s$^2$4p$^4$($^1$D)4d	   &  $^2$D$  _{5/2}$	& 19.4600  & 19.6979 & 19.7968 & 19.8118 & 19.8540 &  19.8612  & 19.7295  &  20.0895 \\
 24  &  4s$^2$4p$^4$($^3$P)4d	   &  $^2$P$  _{3/2}$	& 19.4600  & 19.6900 & 19.7435 & 19.7558 & 19.7939 &  19.8452  & 19.7326  &  20.0904 \\
 25  &  4s$^2$4p$^4$($^1$D)4d	   &  $^2$S$  _{1/2}$	& 19.4600  & 19.7493 & 19.8213 & 19.8254 & 19.8568 &  19.8619  & 19.7807  &  20.1577 \\
 26  &  4s$^2$4p$^4$($^3$P)4d	   &  $^2$D$  _{5/2}$	&	   & 19.8817 & 20.0521 & 20.0682 & 20.1086 &  20.1156  & 19.9539  &  20.2864 \\
 27  &  4s$^2$4p$^4$($^1$D)4d	   &  $^2$F$  _{7/2}$	&	   & 20.1821 & 20.3465 & 20.3623 & 20.4017 &  20.4084  & 20.2493  &  20.5752 \\
 28  &  4s$^2$4p$^4$($^1$D)4d	   &  $^2$D$  _{3/2}$	&	   & 21.3432 & 21.3785 & 21.3886 & 21.4272 &  21.4334  & 21.3805  &  21.7333 \\
 29  &  4s$^2$4p$^4$($^1$D)4d	   &  $^2$P$  _{1/2}$	&	   & 21.8621 & 21.8126 & 21.8191 & 21.8548 &  21.8603  & 21.8929  &  22.2727 \\
 30  &  4s$^2$4p$^4$($^3$P)4d	   &  $^2$D$  _{3/2}$	&	   & 26.3393 & 26.2768 & 26.2813 & 26.3417 &  26.3481  &	  &  26.6524 \\
 31  &  4s$^2$4p$^4$($^1$S)4d	   &  $^2$D$  _{5/2}$	&	   & 26.8104 & 26.8802 & 26.8922 & 26.9603 &  26.9671  &	  &  27.0682 \\
 32  &  3d$^9$4s$^2$4p$^6$         &  $^2$D$  _{5/2}$   & 120.540  & 120.763 & 121.153 & 121.046 & 120.873 &  120.885  & 120.611  &  121.121 \\
 33  &  3d$^9$4s$^2$4p$^6$         &  $^2$D$  _{3/2}$   & 125.110  & 125.522 & 125.884 & 125.811 & 125.638 &  125.650  & 125.389  &  125.921 \\
& & & & & & & & \\ \hline            								                	 
\end{tabular}   								   					       
			      							   					       
\begin{flushleft}													       
{\small
NIST: http://physics.nist.gov/PhysRefData/ASD/levels\_form.html \\
GRASP1: present calculations from the {\sc grasp} code with 63 levels \\ 
GRASP2: present calculations from the {\sc grasp} code with 3490 levels\\
GRASP3: present calculations from the {\sc grasp} code with 4128 levels\\ 
FAC1: present calculations from the {\sc fac} code with 4128 levels  \\  
FAC2: present calculations from the {\sc fac} code with 11,525 levels \\ 										
RELAC: Fournier \cite{kbf} \\ 
GRASP4: AJM  \cite{ajm} \\ 

}															       
\end{flushleft} 

\newpage
\clearpage
\begin{flushleft}
Table 3. Comparison of oscillator strengths (f- values) for some transitions of W XL. $a{\pm}b \equiv a{\times}$10$^{{\pm}b}$.
\end{flushleft}
\begin{tabular}{rllrrrrrrrrrr} \hline
I & J       &  GRASP1  &  GRASP2 &  GRASP3 &  FAC1   &  FAC2  & GRASP4 &   RELAC     & R3    & R4    \\
\hline	      
    1  &  3 & 2.724-3 & 3.036-3 & 3.047-3 & 3.071-3 &  3.073-3 &  2.780-3 & 2.903-3  & 7.3-1 & 9.2-1 \\
    1  &  4 & 1.929-3 & 2.209-3 & 2.255-3 & 2.302-3 &  2.315-3 &  1.830-3 & 2.377-3  & 7.4-1 & 9.3-1 \\
    1  &  5 & 6.938-3 & 6.551-3 & 6.545-3 & 6.453-3 &  6.450-3 &  5.600-3 & 7.760-3  & 8.0-1 & 9.5-1 \\
    1  &  7 & 2.236-4 & 2.955-4 & 3.027-4 & 3.161-4 &  3.166-4 &  2.130-4 & 1.148-4  & 8.7-1 & 1.1-0 \\
    1  &  9 & 3.192-3 & 3.265-3 & 3.178-3 & 2.786-3 &  2.748-3 &  1.690-3 & 3.865-3  & 1.0-0 & 1.1-0 \\
    1  & 11 & 5.333-6 & 2.949-3 & 3.228-3 & 3.169-3 &  3.105-3 &  1.440-3 & 3.798-5  & 9.0-1 & 1.0-0 \\
    1  & 12 & 2.891-1 & 2.636-1 & 2.652-1 & 2.670-1 &  2.673-1 &  2.590-1 & 3.018-1  & 7.9-1 & 9.5-1 \\
    1  & 13 & 5.990-1 & 5.156-1 & 5.187-1 & 5.225-1 &  5.232-1 &  5.130-1 & 5.633-1  & 8.0-1 & 9.6-1 \\
    1  & 14 & 1.887-1 & 1.499-1 & 1.523-1 & 1.563-1 &  1.567-1 &  1.620-1 & 1.585-1  & 7.2-1 & 9.2-1 \\
    1  & 15 & 1.365-3 & 1.282-3 & 1.289-3 & 1.272-3 &  1.269-3 &  1.240-3 & .......  & 7.5-1 & 9.1-1 \\
    1  & 16 & 1.757-4 & 1.228-4 & 1.340-4 & 1.358-4 &  1.380-4 &  1.010-4 & .......  & 9.3-1 & 1.1-0 \\
    1  & 17 & 2.684-2 & 2.630-2 & 2.683-2 & 2.657-2 &  2.662-2 &  2.510-2 & 2.885-2  & 8.4-1 & 9.7-1 \\
    1  & 20 & 7.276-2 & 2.343-1 & 2.461-1 & 2.617-1 &  2.639-1 &  6.630-2 & 9.385-2  & 8.4-1 & 9.5-1 \\
    1  & 22 & 2.747-3 & 9.084-1 & 9.173-1 & 9.251-1 &  9.262-1 &  2.860-3 & 1.018-0  & 8.4-1 & 9.7-1 \\
    1  & 23 & 1.063-0 & 2.928-2 & 2.784-2 & 2.029-2 &  1.993-2 &  9.350-1 & 1.210-2  & 8.6-1 & 9.7-1 \\
    1  & 24 & 5.970-1 & 3.250-1 & 3.158-1 & 3.011-1 &  2.993-1 &  5.380-1 & 5.843-1  & 8.4-1 & 9.6-1 \\
    1  & 25 & 5.270-1 & 4.255-1 & 4.261-1 & 4.232-1 &  4.231-1 &  4.430-1 & 4.730-1  & 8.1-1 & 9.5-1 \\
    1  & 26 & 1.511-1 & 3.238-2 & 2.747-2 & 2.646-2 &  2.613-2 &  1.130-1 & 9.165-2  & 8.3-1 & 9.7-1 \\
    1  & 28 & 4.291-2 & 3.211-2 & 3.220-2 & 3.171-2 &  3.168-2 &  3.540-2 & .......  & 8.5-1 & 9.7-1 \\
    1  & 29 & 6.720-5 & 2.514-4 & 2.096-4 & 1.350-4 &  1.268-4 &  1.680-4 & .......  & 1.8-1 & 1.6-0 \\
    1  & 30 & 8.904-5 & 4.239-6 & 1.945-6 & ....... &  ....... &  2.020-6 & .......  & 9.7-1 & 1.1-0 \\
    1  & 31 & 3.429-4 & 5.361-4 & 5.839-4 & 5.745-4 &  5.716-4 &  5.860-4 & .......  & 7.7-1 & 9.8-1 \\
    2  &  3 & 2.659-6 & 6.999-5 & 7.300-5 & 7.449-5 &  7.495-5 &  5.590-5 & .......  & 5.9-1 & 5.7-1 \\
    2  &  5 & 7.981-4 & 1.234-3 & 1.256-3 & 1.310-3 &  1.316-3 &  1.350-3 & .......  & 2.7-1 & 7.1-1 \\
    2  &  7 & 1.067-4 & 1.124-4 & 1.083-4 & 1.055-4 &  1.046-4 &  1.280-4 & .......  & 1.8-0 & 1.5-0 \\
    2  &  9 & 5.376-3 & 3.899-3 & 4.044-3 & 4.266-3 &  4.300-3 &  5.080-3 & .......  & 2.8-1 & 7.7-1 \\
    2  & 12 & 6.521-7 & 1.564-5 & 1.389-5 & ....... &  ....... &  ....... & .......  & 6.0-2 & 2.7-2 \\
    2  & 14 & 5.442-2 & 4.264-2 & 4.328-2 & 4.367-2 &  4.375-2 &  4.570-2 & .......  & 4.3-1 & 8.1-1 \\
    2  & 15 & 2.594-3 & 1.791-3 & 1.810-3 & 1.815-3 &  1.818-3 &  1.680-3 & .......  & 6.3-1 & 8.8-1 \\
    2  & 16 & 4.335-3 & 3.102-3 & 3.141-3 & 3.207-3 &  3.218-3 &  2.870-3 & 3.053-3  & 6.7-1 & 8.9-1 \\
    2  & 20 & 1.240-3 & 2.919-5 & 4.681-5 & ....... &	.......&  8.350-4 & ....... .& 8.7-1 & 9.9-1 \\
    2  & 24 & 8.250-3 & 1.096-2 & 1.104-2 & 1.086-2 &  1.085-2 &  9.840-3 & .......  & 8.1-1 & 9.3-1 \\
    2  & 25 & 7.863-3 & 3.842-4 & 3.773-4 & 3.652-4 &  3.651-4 &  2.700-3 & .......  & 3.9-3 & 7.4-1 \\
    2  & 28 & 9.637-1 & 8.119-1 & 8.160-1 & 8.216-1 &  8.227-1 &  8.090-1 & 9.135-1  & 7.9-1 & 9.4-1 \\
    2  & 29 & 8.906-1 & 7.384-1 & 7.398-1 & 7.402-1 &  7.406-1 &  7.700-1 & .......  & 7.8-1 & 9.4-1 \\
    2  & 30 & 1.330-0 & 1.052-0 & 1.053-0 & 1.052-0 &  1.053-0 &  1.150-0 & .......  & 8.4-1 & 9.6-1 \\
\hline            								                	 
\end{tabular}   								   					       
			      							   					       
\begin{flushleft}													       
{\small
GRASP1: present calculations from the {\sc grasp} code with 63 levels \\ 
GRASP2: present calculations from the {\sc grasp} code with 3490 levels\\
GRASP3: present calculations from the {\sc grasp} code with 4128 levels\\ 
FAC1: present calculations from the {\sc fac} code with 4128 levels  \\  
FAC2: present calculations from the {\sc fac} code with 11,525 levels \\ 										
GRASP4: Aggarwal et al, Can J. Phys. 91 (2013) 394 \\
RELAC: Fournier, ADNDT 68 (1998) 1 \\
R3: ratio of velocity and lengths f- values corresponding to the GRASP3 calculations \\
R4: ratio of velocity and lengths f- values corresponding to the GRASP4 calculations \\
															       
}															       
\end{flushleft} 


\newpage
\clearpage
\begin{flushleft}
Table 4. Comparison of lifetimes ($\tau$, s)  for some levels of W XL. $a{\pm}b \equiv a{\times}$10$^{{\pm}b}$.
\end{flushleft}
\begin{tabular}{rllrrrlrrrrrr} \hline
 & & & & & & & &  \\
Index  & Configuration      & Level                    &  GRASP3    &  GRASP4a &  GRASP4b  & GRASP4a (Dominant A- values, s$^{-1}$)  \\
& & & & & & & &   \\ \hline
& & & & & & & &   \\
  1  &  4s$^2$4p$^5$	   	   &  $^2$P$^o_{3/2}$  & .......    & .......  & .......   & 	.......				    \\
  2  &  4s$^2$4p$^5$	   	   &  $^2$P$^o_{1/2}$  & 1.314-07   & 1.312-07 & 1.37-07   &  1--2 M1=7.29+06			    \\
  3  &  4s$^2$4p$^4$($^3$P)4d	   &  $^4$D$  _{3/2}$  & 3.218-10   & 3.422-10 & 3.43-10   &  1--3 E1=2.92+09			    \\
  4  &  4s$^2$4p$^4$($^3$P)4d	   &  $^4$D$  _{5/2}$  & 6.346-10   & 7.564-10 & 7.57-10   &  1--4 E1=1.32+09			    \\
  5  &  4s$^2$4p$^4$($^3$P)4d	   &  $^4$P$  _{1/2}$  & 7.170-11   & 8.083-11 & 8.24-11   &  1--5 E1=1.21+10			    \\
  6  &  4s$^2$4p$^4$($^3$P)4d	   &  $^4$F$  _{7/2}$  & 4.765-03   & 4.119-03 & 6.07-03   &  1--6 M2=1.64+02, 4--6 M1=7.84+01      \\
  7  &  4s$^2$4p$^4$($^1$S)4d	   &  $^2$D$  _{3/2}$  & 2.692-09   & 3.595-09 & 3.82-09   &  1--7 E1=2.62+08			    \\
  8  &  4s$^2$4p$^4$($^3$P)4d	   &  $^2$F$  _{7/2}$  & 2.238-05   & 2.209-05 & 8.00-05   &  1--8 M2=1.25+04, 4--8 M1=2.44+04      \\
  9  &  4s$^2$4p$^4$($^3$P)4d	   &  $^2$P$  _{1/2}$  & 1.046-10   & 1.633-10 & 2.18-10   &  1--9 E1=4.56+09, 2--9 E1=1.56+09      \\
 10  &  4s$^2$4p$^4$($^3$P)4d	   &  $^4$F$  _{9/2}$  & 3.703-05   & 3.731-05 & 1.27-01   &  6--10 M1=2.68+04  		    \\
 11  &  4s$^2$4p$^4$($^3$P)4d	   &  $^4$P$  _{5/2}$  & 3.064-10   & 6.624-10 & 6.63-10   &  1--11 E1=1.51+09  		    \\
 12  &  4s$^2$4p$^4$($^3$P)4d	   &  $^4$P$  _{3/2}$  & 2.363-12   & 2.326-12 & 2.33-12   &  1--12 E1=4.30+11  		    \\
 13  &  4s$^2$4p$^4$($^1$D)4d	   &  $^2$F$  _{5/2}$  & 1.755-12   & 1.705-12 & 1.70-12   &  1--13 E1=5.87+11  		    \\
 14  &  4s4p$^6$                   &  $^2$S$  _{1/2}$  & 1.658-12   & 1.545-12 & 1.62-12   &  1--14 E1=6.20+11  		    \\
 15  &  4s$^2$4p$^4$($^3$P)4d	   &  $^4$D$  _{1/2}$  & 1.221-10   & 1.249-10 & 1.58-10   &  1--15 E1=6.35+09, 2--15 E1=1.66+09    \\
 16  &  4s$^2$4p$^4$($^3$P)4d	   &  $^4$F$  _{3/2}$  & 5.133-10   & 5.615-10 & 3.73-09   &  2--16 E1=1.51+09  		    \\
 17  &  4s$^2$4p$^4$($^3$P)4d	   &  $^4$F$  _{5/2}$  & 2.070-11   & 2.165-11 & 2.16-11   &  1--17 E1= 4.62+10 		    \\
 18  &  4s$^2$4p$^4$($^1$D)4d	   &  $^2$G$  _{7/2}$  & 2.604-07   & 2.554-07 & 4.74-04   &  6--18 M1=3.03+6			    \\
 19  &  4s$^2$4p$^4$($^3$P)4d	   &  $^4$D$  _{7/2}$  & 1.356-07   & 1.349-07 & 7.91-05   &  8--19 M1=2.09+06, 10--19 M1=2.99+06   \\
 20  &  4s$^2$4p$^4$($^1$D)4d	   &  $^2$P$  _{3/2}$  & 1.317-12   & 4.736-12 & 4.77-12   &  1--20 E1=2.11+11  		    \\
 21  &  4s$^2$4p$^4$($^1$D)4d	   &  $^2$G$  _{9/2}$  & 2.445-07   & 2.404-07 & 8.39-05   &  10--21 M1=3.10+06 		    \\
 22  &  4s$^2$4p$^4$($^3$P)4d	   &  $^2$F$  _{5/2}$  & 1.732-11   & 1.622-10 & 1.63-10   &  1--22 E1=6.16+09  		    \\
 23  &  4s$^2$4p$^4$($^1$D)4d	   &  $^2$D$  _{5/2}$  & 5.219-13   & 4.949-13 & 4.95-13   &  1--23 E1=2.02+12  		    \\
 24  &  4s$^2$4p$^4$($^3$P)4d	   &  $^2$P$  _{3/2}$  & 1.003-12   & 5.724-13 & 5.74-13   &  1--24 E1=1.74+12  		    \\
 25  &  4s$^2$4p$^4$($^1$D)4d	   &  $^2$S$  _{1/2}$  & 3.716-13   & 3.452-13 & 3.46-13   &  1--25 E1=2.89+12  		    \\
 26  &  4s$^2$4p$^4$($^3$P)4d	   &  $^2$D$  _{5/2}$  & 1.665-11   & 4.050-12 & 4.01-12   &  1--26 E1=2.47+11  		    \\
 27  &  4s$^2$4p$^4$($^1$D)4d	   &  $^2$F$  _{7/2}$  & 2.046-07   & 2.018-07 & 6.41-04   &  10--27 M1=2.69+06 		    \\
 28  &  4s$^2$4p$^4$($^1$D)4d	   &  $^2$D$  _{3/2}$  & 1.226-12   & 1.165-12 & 7.44-12   &  2--28 E1=7.24+11  		    \\
 29  &  4s$^2$4p$^4$($^1$D)4d	   &  $^2$P$  _{1/2}$  & 7.449-13   & 6.751-13 & 7.47-10   &  2--29 E1=1.48+12  		    \\
 30  &  4s$^2$4p$^4$($^3$P)4d	   &  $^2$D$  _{3/2}$  & 6.228-13   & 5.512-13 & 8.66-08   &  2--30 E1=1.81+12  		    \\
 31  &  4s$^2$4p$^4$($^1$S)4d	   &  $^2$D$  _{5/2}$  & 4.386-10   & 4.305-10 & 4.33-10   &  1--31 E1=2.30+09  		    \\
 32  &  3d$^9$4s$^2$4p$^6$         &  $^2$D$  _{5/2}$  & 2.310-13   & 2.797-13 & 2.79-13   &  1--32 E1=3.55+12  		    \\
 33  &  3d$^9$4s$^2$4p$^6$         &  $^2$D$  _{3/2}$  & 1.857-13   & 2.071-13 & 2.06-13   &  2--33 E1=4.25+12  		    \\
& & & & & & & & \\ \hline            								                	 
\end{tabular}   								   					       
			      							   					       
\begin{flushleft}													       
{\small
GRASP3: present calculations from the {\sc grasp} code with 4128 levels\\ 
GRASP4a: present calculations from the {\sc grasp} code with 638 levels\\ 										
GRASP4b: AJM \cite{ajm} \\ 
															       
}															       
\end{flushleft} 
\end{document}